# On the extension of Newton's second law to theories of gravitation in curved space-time

M. ARMINJON (GRENOBLE)

**Abstract-** We investigate the possibility of extending Newton's second law to the general framework of theories in which special relativity is locally valid, and in which gravitation changes the flat Galilean space-time metric into a curved metric. This framework is first recalled, underlining the possibility to uniquely define a space metric and a local time in any given reference frame, hence to define velocity and momentum in terms of the local space and time standards. It is shown that a unique consistent definition can be given for the derivative of a vector (the momentum) along a trajectory. Then the possible form of the gravitation force is investigated. It is shown that, if the motion of free particles has to follow space-time geodesics, then the expression for the gravity acceleration is determined uniquely. It depends on the variation of the metric with space and time, and it involves the velocity of the particle.

## 1. Introduction

This work comes from an attempt to explore the possibility of extending the "logic of absolute motion", which prevails in the Lorentz-Poincaré interpretation of special relativity [8-9, 15, 20-24], so as to obtain a consistent theory of gravitation. Thus, a theory with a preferred frame has been tentatively proposed [1-4]. Just like general relativity (GR), this theory endows the space-time with a curved metric. Just like in GR, special relativity (SR) holds true locally in this tentative theory. However, an extension of Newton's second law, or rather of its modified expression valid in SR, has been defined for a test particle (mass point or photon) in the most general situation within this investigated theory [4]. As it will be reported here, the way used in this theory to define Newton's second law in a "curved space-time" turns out to be both natural and general in its principle. Hence, it has been tried to find in the literature such a natural and general extension, but this quest has not been really successful. Apart from approximate equations occurring in "post-Newtonian" treatments, two exact extensions of Newton's second law to relativistic theories of gravitation can be found among well-known textbooks: Landau & Lifchitz [11, §88] define this law for a *constant* gravitation field, and Møller [18, §110] "tries to write [the equations of space-time geodesics] in the form of three-dimensional vector equations" in a general case but, as his sentence suggests, and as will be discussed below (note [1] and Sect. 4), his attempt is not fully satisfactory. Jantzen *et al*. [10] review and unify the various attempts, including the important work of Cattaneo [6-7], to "split space-time into space plus time" and to rewrite the relativistic equations of motion with "spatial gravitational forces". It appears from their review that three different definitions have been introduced, by various authors, for the time-derivative of the momentum. These definitions will be examined in Sect. 4. It will appear that one does not obey Leibniz' rule, while none of the other two does involve *only* the *separate* ingredients "space metric" and "time metric" in a given reference frame- as should be true for a natural extension of Newton's second law. However, it seems that one has good reasons to search for such extension and hence to find this "missing link" [17] between classical and relativistic mechanics.

Indeed, the Lorentz-Poincaré construction of special relativity [15, 20-21], fully developed by Jánossy [8-9] and Prokhovnik [22-24], obtains the "relativistic" effects as being all consequences of the "true" Lorentz contraction assumed to affect all bodies in motion with respect to the "ether". As it has been recently reestablished [27] against contrary statements, it is impossible to consistently measure the anisotropy in the *one-way* velocity of light. This makes the Lorentz-Poincaré version empirically undistinguishable from the Einstein version of SR [22]. The Lorentz-Poincaré interpretation allows to concile special relativity with our intuitive notion of distinct space and time, and thus with the most crucial concepts of classical mechanics. However, special relativity does not describe gravitation: for gravitation, general relativity is the current tool. But in GR, the laws of motion become a consequence of the space-time curvature, e.g. the "free" particles are assumed to follow the geodesic lines of the space-time metric. Thus, at least as long as the geodesic formulation of motion has not been derived from a generalization of Newton's second law, one is enforced to give a physical status to space-time in GR. On the other hand, despite the experimental success of GR, it remains unsolved problems as regards gravitation. We may mention the problem of the singularity occurring with the gravitational collapse of very massive objects, and the need to postulate "dark matter" in order to explain stellar motion in galaxies. We should also mention the questions on the influence of the coordinate condition in GR, which were raised a long time ago (e.g. Papapetrou [19]), but that have been newly discussed by Logunov *et al.* [13-14]. Logunov *et al.* present detailed arguments against the usual agreement that, in GR, the choice of the coordinate condition has no physical consequence. It thus may be worth to investigate alternative, speculative theories and to ask questions on the formulation of motion.

In this paper, an extension of Newton's second law will be given for theories of gravitation in curved space-time in which SR is locally valid, *including GR*. In doing so, care will be taken to maintain *space* covariance in a given reference frame, in order that the force be properly defined. However, no attempt will be made to investigate the transformation of the force from one reference frame to another. Section 2 will be focused on the definition of the right-hand side of Newton's law, i.e. the time-derivative of the momentum: it will be shown that this may be defined from rather compelling principles, up to the same parameter $\lambda$ as in the tentative theory [4], and which also must be $\lambda = 1/2$ if Leibniz' rule is to apply. In Section 3, it will be investigated which form of the gravitation force is compatible with Einstein motion (for "free" particles), i.e. motion along space-time geodesics. In a first step, Leibniz' rule will not be imposed but it will be assumed, in analogy with Newtonian theory, that the gravitation force depends linearly on the spatial derivatives of the metric and does not depend on its time-derivative. In a second step, Leibniz' rule will be assumed, but no restriction to the gravitation force will be imposed. In Section 4, the three anterior definitions of the time-derivative of a spatial vector, reviewed by Jantzen *et al.*, will be examined from the point of view of "consistency" (validity of Leibniz' rule), and "naturalness" (space plus time separation).

**2. Definition of Newton's second law for a (pseudo-) Riemannian space-time metric**

2.1 *Some clarification on the kind of theories considered*

We suppose that, according to some gravitation theory, the physical standards of space and time are influenced by a gravitation field, but that SR holds true locally (GR is the prototype of such gravitation theories, of course). It will be useful to recall in some detail what is meant by this, not the least because it will make clear that this framework does not preclude to consider a preferred-frame theory, nor does this framework imply that a fundamental physical meaning must be given to the mathematical concept of space-time. It will also give the way to separate the force into a gravitational force or rather a mass force, and a non-gravitational force.

(**i**) According to a theory of this kind, our space and time measurements may be arranged so as to be described by a metric **γ** with (1,3) signature on a 4-dimensional, "space-time" manifold. This may be done as follows. Any possible *reference frame* F, physically defined by a *spatial network of "observers"* (each one equipped with a ruler and a clock, all made in the same factory, say), allows one to define (in many ways, actually) an associated coordinate system $(x^\alpha)$ ($\alpha = 0,...,3$), with $x^0$ the time coordinate and $x^i$ ($i = 1, 2, 3$) the space coordinates, so that *each observer has constant space coordinates*. Moreover, $t = x^0/c$ is the "formal date" assigned to an event occurring at a point specified by the space coordinates $x^i$ ($t$ has in general no immediate relation to real time-measurements made by the observer at this point). The observers in the same frame F are not necessarily at rest with each other, i.e. they may find that their mutual distances are not conserved (case of a deformable frame). The manifold structure of the space-time means simply that the same physical events will be given different space and time coordinates by different networks of observers, say $(x^\alpha)$ and $(x'^\alpha)$, and that the correspondence between $(x^\alpha)$ and $(x'^\alpha)$ is locally smooth (for smoothly deforming networks). So we have a space-time manifold $M^4$. The elements (points) of the spatial network cannot be identified with *points* in that manifold but with "world lines", thus with *lines* in space-time. Hence, from the point of view of "space-time", a reference frame is a 3-D differentiable manifold N whose each point is a (time-like) differentiable mapping from the real line into the space-time $M^4$, moreover N is diffeomorphic to any spatial section of $M^4$ (this is only the sketch of a rigorous definition; from the point of view of "space + time", a much simpler definition may be proposed [1]). Note that many new coordinate systems $(x'^\alpha)$ do *not* change the reference frame (network) specified by one system $(x^\alpha)$: the frame remains unaltered if and only if the change of the space coordinates does not depend on the time coordinate, i.e. $\partial x'^i/\partial x^0 = 0$. Up to this point, it seems that no physically restrictive assumption is involved (except, of course, for the fact that "classical" physics, not quantum physics, is envisaged here).

The assumption that SR applies locally is the one which allows to define a (1,3) space-time metric. This assumption means, in the first place, this: in any reference frame, the velocity of light, as measured on a to-and-fro path between infinitesimally distant positions, is always the same constant $c$. Under this condition, the link between physical space and time measurements and the metric $\gamma$ may be described as in Landau & Lifchitz [11], it is based on using the Poincaré-Einstein synchronization convention for infinitesimally distant clocks. Thus the proper time along the trajectory of a mass point ("time-like" line in space-time), i.e. the time $\tau$ measured by a clock bound to the moving point, is directly given by metric $\gamma$:

$$(2.1) \qquad ds^2 = c^2 \, d\tau^2 = \gamma_{\alpha\beta} \, dx^\alpha \, dx^\beta.$$

Also, the distance $dl$ between neighbouring observers (of a given frame F, specified by a coordinate system), as they find by using their rulers, or by measuring the interval $d\tau$ of their proper time that it takes for a light signal to go forth and back, is expressed by a space metric tensor $\mathbf{h} = \mathbf{h}_F$ (it depends on the frame F):

$$(2.2) \qquad dl^2 = (c \, d\tau/2)^2 = h_{ij} \, dx^i \, dx^j, \quad h_{ij} = -\gamma_{ij} + (\gamma_{0i} \, \gamma_{0j} / \gamma_{00}).$$

Moreover, a synchronized local time $t_\mathbf{x}(\xi)$ may be defined along any open line in space-time [i.e. a piecewise differentiable and *one-to-one* mapping $\xi \to (x^\alpha(\xi))$ defined on a closed segment of the real line], such that its variation along the given trajectory is given by:

$$(2.3) \qquad \frac{dt_\mathbf{x}}{d\xi} = \frac{\sqrt{\gamma_{00}}}{c} \left( \frac{dx^0}{d\xi} + \frac{\gamma_{0i}}{\gamma_{00}} \frac{dx^i}{d\xi} \right).$$

As emphasized by Cattaneo [6], the interval $dt_\mathbf{x}$ is invariant under any coordinate transformation that leaves the reference frame unchanged ("internal transformation") and has thus an objective physical meaning. If the $\gamma_{0i}$ components ($i = 1, 2, 3$) are identically equal to zero, the synchronization convention implies that events occuring at a given value of $x^0$ are simultaneous in the frame F, independently of their spatial coordinates (this may be seen on Eq. (2.3)). Hence $x^0$ is a "universal time" in the frame F. As a consequence, if one uses such coordinates ($x^\alpha$), then the trajectory of any test particle may always be parametrized with the coordinate time $t$ itself and moreover the local time has the simple expression

$$dt_\mathbf{x}/dt = \sqrt{\gamma_{00}} \equiv \beta \qquad (2.4)$$

The expression (2.4) of the local time has the immediate physical meaning of showing how clocks are affected by the gravitation field (usually they are slowed down, i.e. $\gamma_{00}$ decreases towards the gravitational attraction). The property "$\gamma_{0i} = 0$" holds true after any coordinate transformation of the form $x'^0 = \phi(x^0)$, $x'^i = \psi^i(x^1, x^2, x^3)$. Thus it is indeed a characteristic of a given frame F. The restriction to space-independent transformation of time, $x'^0 = \phi(x^0)$, reflects simply the global synchronization. Using this time transformation, one may impose that the local time at a given point bound to the frame, $\mathbf{x}_0 = (x_0^i)$, coincides with the universal time [i.e. $\gamma_{00}(x^0, (x_0^i)) = 1 \quad \forall x^0$], and then only a shift of $x^0$ is left free. The $\gamma_{00}$ component is invariant under the remaining, purely spatial coordinate changes.

(**ii**) The other assumption involved, in saying that SR applies locally, is that the laws of non-gravitational physics are "formally unaffected" by gravitation, in the following sense: in the absence of gravitation, any such law must (or should) be formulated in the frame of SR. Then, in the absence of gravitation, it may be expressed in a generally covariant form, in replacing the partial derivatives, valid in Galilean coordinates, by the covariant derivatives with respect to the *flat* space-time metric $\gamma^0$ [Galilean coordinates are ones in which the flat metric $\gamma^0$ has the canonical diagonal form, $\gamma^0_{\mu\nu} = \eta_{\mu\nu}$ with $(\eta_{\mu\nu}) \equiv \text{diag}(1, -1, -1, -1)$]. Now the *assumption* is that, in the presence of gravitation and hence (according to a theory of the class considered here) with a *curved* metric $\gamma$, *the expression of any such law is extended to this situation simply in substituting $\gamma$ for $\gamma^0$*. This assumption is quite natural: physics must be described in terms of the local space and time standards which (cf. point (**i**)) are ruled by metric $\gamma$ in the frame of SR. And at the local or rather at the infinitesimal scale, the presence or absence of curvature plays little or no rôle, i.e. any metric behaves (in many respects though not in all) as a flat metric in the infinitesimal. Some ambiguity may yet arise when trying to use this assumption, if differential expressions of order greater than one are involved: since Schwarz' theorem does not apply to covariant derivatives for a curved metric, different higher-order expressions may become identical for a flat metric and yet remain distinct for a curved one (e.g. Will [26]). In a such case, a comparison with experiment may either decide between the possibilities, or show that they do not differ significantly. Such empirical procedure might lead, of course, to different choices for different gravitation theories i.e. for different metrics **γ** in the same physical situation, and thus could create a biase when testing alternative theories.

### 2.2 *Extended Newton law for a constant gravitation field*

Let us first consider the *static* case, i.e. the case where a frame F exists, defined by a coordinate system $(x^\alpha)$, in which all components $\gamma_{\alpha\beta}$ of metric **γ** are independent of $x^0$, and moreover the $\gamma_{0i}$ ($i = 1, 2, 3$) components are zero. The first property holds true after any coordinate transformation of the form $x'^0 = a\, x^0 + \phi(x^1, x^2, x^3)$, $x'^i = \psi^i(x^1, x^2, x^3)$, thus a different range for the time transformation than for the second property, discussed above. Then, the *right-hand side* of Newton's second law, valid for SR, i.e. $d\mathbf{P}/dt$ with **P** the momentum including the velocity-dependent mass, is easy to extend to any such theory of gravitation. The velocity **v** of a test particle (relative to the frame F) is measured with the local time $t_\mathbf{x}$ of the momentarily coincident observer in the frame F, and its modulus $v$ is defined with the point-dependent (Riemannian) space metric **h** in the frame F. Thus

(2.5) $\qquad\qquad v^i \equiv dx^i/dt_\mathbf{x}\,, \qquad v \equiv [\mathbf{h}(\mathbf{v},\mathbf{v})]^{1/2} = (h_{ij}\, v^i v^j)^{1/2}.$

The momentum is hence for a time-like test particle (mass point):

(2.6) $\qquad\qquad \mathbf{P} \equiv m(v)\,\mathbf{v}, \quad m(v) \equiv m(v{=}0).\gamma_v \equiv m(0).(1-v^2/c^2)^{-1/2}$

(using the mass-velocity relation of SR)[1]. For a light-like test particle (photon), one substitutes the mass content of the energy for the inertial mass $m(v)$. Then we must define the derivative of the momentum with respect to the local time. Thus in general we have to define the derivative of a vector $\mathbf{w} = \mathbf{w}(\chi)$ attached to a point $\mathbf{x}(\chi) = (x^i(\chi))$ which moves, as a function of the real parameter $\chi$, in some Riemannian space : here this space is the 3-D domain $N = N_F$ constituted by the spatial network which defines the considered frame F. Hence the points in N are specified by their constant space coordinates $x^i$, $i = 1,2,3$, and N is equipped with the space metric $\mathbf{h}$. The derivative must be defined as the "absolute" derivative (e.g. Brillouin [5], Lichnerowicz [12]), which is a space vector and accounts for the (merely spatial) variation of the space metric along the trajectory:

$$(2.7) \qquad \left(\frac{D\mathbf{w}}{D\chi}\right)^i \equiv \frac{dw^i}{d\chi} + \Gamma^i_{jk}\, w^j\, \frac{dx^k}{d\chi},$$

where the $\Gamma^i_{jk}$'s are the Christoffel symbols of metric $\mathbf{h}$ in coordinates $(x^i)$. As shown in ref. [2], the use of Eq. (2.7) is *enforced* if one wants that Leibniz' rule applies and that the derivative cancels for a vector $\mathbf{w}$ that is parallel-transported (relative to the space metric $\mathbf{h}$) along the trajectory. This is considered important, because it means that Eq. (2.7) is not merely one possible formal rule to obtain a space-contravariant vector, but the unique consistent definition for the time-derivative of a vector along a trajectory, in the case of a time-independent metric. Now the *left-hand side* of Newton's second law is just the force. This may be decomposed into a "non-gravitational" force $\mathbf{F}_0$, which should have the same expression for any gravitation theory in the considered class[2], and a "gravitational" force $\mathbf{F}_g$ whose expression, of course, will depend on the theory. Note that $\mathbf{F}_g$ *will generally contain "inertial" forces as well* (since a general reference frame is considered here), hence "mass force" would be a more appropriate denomination [1]. Thus finally:

$$(2.8) \qquad \mathbf{F}_0 + \mathbf{F}_g = D\mathbf{P}/Dt_\mathbf{x}.$$

---

[1] Equation (2.6) implicitly assumes that the rest mass $m(0)$ is the same constant $m_0$, independently of the gravitation field. This may be seen as an immediate consequence of defining the inertial mass $m$ as the ratio **P**/**v** ($=P^i/v^i$) and assuming that the $P^i$'s are the spatial components of the 4-momentum, this being in turn assumed to have the form $P^\alpha = m_0\, dx^\alpha/d\tau$ with a constant $m_0$. This is consistent with Landau & Lifchitz [11]. On the other hand, Møller [18] defines the inertial mass as the ratio $m' = \mathbf{P}/\mathbf{v}_0$ with $\mathbf{v}_0 = d\mathbf{x}/dt$, thus $m' = m\, dt/dt_\mathbf{x}$, hence his rest mass $m'_0 = m'(\mathbf{v}_0 = 0) = m_0\, dt/dt_\mathbf{x}$ depends on the gravitation field. However, the definition of $\mathbf{v}_0$ and hence that of $m'_0$ depend on the chosen time coordinate $t$ even in a given frame, while the velocity **v** used by Landau & Lifchitz (and used here) depends only on the reference frame, as it should.

---

[2] The expression of $\mathbf{F}_0$ is taken from the situation without gravitation: thus, as recalled in point (**ii**) of Subsect. 2.1, it involves the field $\boldsymbol{\gamma}$ (in the place of the flat metric $\boldsymbol{\gamma}^0$), and it depends on the non-gravitational fields- in practice these are the electromagnetic field and/or thermomechanical fields (the nuclear fields are very microscopic matter fields and moreover their current theory does not belong to classical physics, i.e. their influence cannot be described in terms of deterministic trajectories of mass points). A "*free*" particle is one which crosses a region free from matter and electromagnetic field: for such a particle, the force $\mathbf{F}_0$ will be zero *independently of the reference frame considered*.

Using the same equations (2.3) and (2.5) to (2.7), the same definition may and must be used in the *stationary* case, in which the $\gamma_{\alpha\beta}$'s remain time-independent, but the $\gamma_{0i}$ components may be non-zero: although a synchronized local time cannot be defined in the frame F as a whole if the $\gamma_{0i}$'s are non-zero, what matters is that it is uniquely defined along the trajectory followed by the considered particle (provided that it follows an open line in space-time: a closed line would mean a travel back in time).

### 2.3 *Extended Newton law for a general gravitation field*

In the general case where the gravitation field is not constant in the frame F, the new feature is that now the space-time metric $\gamma$ depends also on $x^0$. Hence also the space metric **h** [Eq. (2.2)] varies, not only as as a function of the space coordinates $x^i$ (as is natural for a general Riemannian metric on a space depending on these coordinates), but also as a function of the time coordinate $x^0$. What is relevant for Newton's second law is, more precisely, the variation of **h** along a trajectory (of a test particle), i.e. the fact that *our spatial network* N *is equipped with a metric field* $\mathbf{h}_\chi$ *that changes as the parameter $\chi$ evolves on the trajectory*, thus for any value of $\chi$ and at every point $X \in$ N we have a covariant tensor $\mathbf{h}_\chi(X)$. In our case, the variation of the metric field with $\chi$ is due to the variation of **h** with the point in space-time, thus in coordinates:

$$h_{\chi\, ij}\,[(x^k)_{k=1,2,3}] \equiv h_{ij}[x^0(\chi),\,(x^k)_{k=1,2,3}].$$

Moreover, we have a preferred parameter $\chi = t_\mathbf{x}$ on the trajectory. It is easy to convince oneself that nothing needs to be changed in Eqs. (2.3), (2.5) and (2.6), because they involve only the local components of the metric (which now become its local and "current" components), not its variation. In order to properly define an extension of (2.7), let us list the properties that should be satisfied by this searched derivative of a vector on a trajectory in a manifold equipped with a variable metric:

(a) It must be a (space) vector, i.e. it must be contravariant for any coordinate transformation of the form $x'^i = x'^i(x^j)$.
(b) It must be linear in **w**. More precisely, it must obviously have the form

$$D\mathbf{w}/D\chi = (d\mathbf{w}/d\chi)_{\chi = \chi_0} + \mathbf{t'}.\mathbf{w}(\chi_0),$$

with $\chi_0$ the point of the trajectory where the derivative is to be calculated, and where $\mathbf{t'}$ is a mixed second-order (space) tensor, transforming a (space) vector into another one.

(c) It must reduce to (2.7) if the metric field $\mathbf{h}_\chi$ does *not* depend on $\chi$.

(d) It should account for the variation of metric $\mathbf{h}_\chi$ as a function of $\chi$.

(e) It must be multiplied by $d\chi/d\zeta$ if $\chi$ is changed to $\zeta = \phi(\chi)$.

(f) It must satisfy Leibniz' derivation rule for the derivative of a scalar product, i.e.

(2.9) $$\frac{d}{d\chi}\left(\mathbf{h}_\chi(\mathbf{w},\mathbf{z})\right) = \mathbf{h}_\chi(\mathbf{w},\frac{D\mathbf{z}}{D\chi}) + \mathbf{h}_\chi(\frac{D\mathbf{w}}{D\chi},\mathbf{z}),$$

in which it is understood that, on the left, the variation of metric $\mathbf{h}$ with $x^0$ is accounted for, as becomes obvious if one writes down explicitly the scalar product:

(2.10) $$\mathbf{h}_\chi(\mathbf{w}, \mathbf{z}) = h_{ij}[(x^\alpha(\chi))_{\alpha = 0,\ldots,3}] \, w^i(\chi) \, z^j(\chi).$$

(Hence, it is likely that (f) implies (d).)

First, we note that definition (2.7) still makes sense, and satisfies requirements (a), (b), (c) and (e). Of course, it is now specified that the Christoffel symbols of metric $\mathbf{h}$ are those at the relevant position *and "time"*, thus in (2.7):

(2.11) $$\Gamma^i_{jk} = \Gamma_\chi{}^i_{jk}[(x^l)_{l=1,2,3}] = \Gamma^i_{jk}[(x^\alpha(\chi))_{\alpha = 0,\ldots,3}].$$

The "candidate" thus defined by Eq. (2.7) will be now denoted by $D_0\mathbf{w}/D\chi$. It does not satisfy (d) [nor (f), in fact], for it amounts in substituting the metric $\mathbf{h}_{\chi_0}$ of the "time" $a = x^0(\chi_0)$ for the variable metric $\mathbf{h}_\chi$. From (b) and (c), it follows that we have to search an expression in the form

(2.12) $$D\mathbf{w}/D\chi = D_0\mathbf{w}/D\chi + \mathbf{t}.\mathbf{w}(\chi_0),$$

in which $\mathbf{t}$ is a mixed second-order space tensor [indeed, the ordinary derivative $d\mathbf{w}/d\chi = (dw^i/d\chi)$ is already involved in $D_0\mathbf{w}/D\chi$, Eq. (2.7)]. But to satisfy (d), it is hence necessary that this tensor involve the variation of metric $\mathbf{h}_\chi$ with $\chi$, due to the variation of $\mathbf{h}$ with $x^0$:

$$\frac{\partial h_{\chi ij}}{\partial \chi} = \frac{\partial h_{ij}}{\partial x^0}\frac{dx^0}{d\chi}.$$

Thus tensor **t** must contain either $h_{ij,0}$ terms or $h^{ij}{}_{,0}$ ones, with $(h^{ij})$ the inverse matrix of $(h_{ij})$. In order to be a mixed tensor and satisfy (e), **t** should have the form

(2.13) $\quad t^i{}_k = h^{ij}\, h_{jk,0}\,(dx^0/d\chi),\quad$ or $\quad t'^i{}_k = h^{ij}{}_{,0}\,(dx^0/d\chi)\, h_{jk},$

or any linear combination of these two tensors. But since $h^{ij}\, h_{jk} = \delta_{ik}$, we have $\mathbf{t} + \mathbf{t}' = 0$, so that, without imposing Leibniz' rule, we are left with a one-parameter family of candidates:

(2.14) $\quad D_\lambda \mathbf{w}/D\chi \equiv D_0 \mathbf{w}/D\chi + \lambda\, \mathbf{t}.\mathbf{w}.$

Finally, nearly the same short calculation as in ref. [4] shows that *Leibniz' rule (2.9) imposes $\lambda = 1/2$*, hence only one definition of the derivative remains:

(2.15) $\quad D\mathbf{w}/D\chi \equiv D_0 \mathbf{w}/D\chi + (1/2)\, \mathbf{t}.\mathbf{w}, \quad \mathbf{t} \equiv \mathbf{h}_\chi{}^{-1}.\frac{\partial \mathbf{h}_\chi}{\partial \chi} \equiv \mathbf{h}^{-1}.\frac{\partial \mathbf{h}}{\partial x^0}\frac{dx^0}{d\chi},$

or in coordinates:

(2.16) $\quad \left(\frac{D\mathbf{w}}{D\chi}\right)^i \equiv \frac{dw^i}{d\chi} + \Gamma^i{}_{jk}\, w^j\, \frac{dx^k}{d\chi} + \frac{1}{2}\, h^{ij}\, h_{jk,0}\, \frac{dx^0}{d\chi}\, w^k.$

Thus, a theory of the kind considered should provide an expression for the mass force $\mathbf{F}_g$, and this expression would depend on what the theory considers as "the gravitation field" (this may include the space-time metric $\boldsymbol{\gamma}$, in any case it must determine $\boldsymbol{\gamma}$). Then one and only one "Newton law" can be consistently stated in such a theory: it is Eq. (2.8), where the momentum **P** is given by Eq. (2.6) and its derivative $D\mathbf{P}/Dt_\mathbf{x}$ is calculated using rule (2.16). The trajectory $\xi \rightarrow (x^\alpha(\xi))$ being defined with the help of an arbitrary parameter $\xi$, the variation of the local time $\chi = t_\mathbf{x}$ along the trajectory is given by Eq. (2.3).

### 2.4 *Comments and link with the investigated preferred-frame theory*

It is seen that the derivative of the momentum is defined in any possible reference frame (and it depends on the frame). Hence, if a theory gives a covariant expression for $\mathbf{F}_g$ and $\boldsymbol{\gamma}$, the extended second Newton law does not restrict the covariance of the theory. On the other hand, a preferred-frame theory may give $\mathbf{F}_g$ and $\boldsymbol{\gamma}$ in one reference frame only; if one were able to calculate the transformation law of the derivative $D\mathbf{P}/Dt_\mathbf{x}$, then this same law would apply to the force, so the law of motion would be reexpressed in a covariant form.

The investigated ether theory [1-4], which is indeed non-covariant, starts from a heuristic interpretation of gravity as Archimedes' thrust in a perfectly fluid "micro-ether" (the rigid ether frame E considered by Lorentz and Poincaré would be defined by the average motion of this "micro-ether" at a very large scale). The transition to account for "relativistic" effects is based on a formulation of Einstein's equivalence principle, natural in this preferred-frame theory: the equivalence is stated to exist between the absolute metric effects of uniform motion and gravitation. This leads to postulate a gravitational contraction (resp. a dilation) of the space (resp. time) standards, depending on the field of the "ether pressure" $p_e$, thus getting a curved (Riemannian) space metric $\mathbf{g}$ and a local time $t_\mathbf{x}$ in the ether frame E, which together build a curved space-time metric $\boldsymbol{\gamma}$ [2-3]. This theory gives $\mathbf{F}_g$ and $\boldsymbol{\gamma}$ in the ether frame E only, as a function of the *scalar* gravitation field $p_e$, or the associated fields $f$ and $\beta$ with

$$(2.17) \qquad f = \beta^2 = (p_e / p_e^\infty)^2 \leq 1,$$

where $p_e^\infty = p_e^\infty(T)$ is the reference pressure (which, for an insular matter distribution, is asymptotically reached at large distance from matter. Here, $T$ is the "absolute time"). The gravitation force is assumed to be

$$(2.18) \qquad \mathbf{F}_g = m(v)\, \mathbf{g},$$

with $\mathbf{g}$ the gravity acceleration, given by

$$(2.19) \qquad \mathbf{g} = -c^2 \frac{\operatorname{grad}_\mathbf{g} p_e}{p_e} = -c^2 \frac{\operatorname{grad}_\mathbf{g} \beta}{\beta} = -\frac{c^2}{2} \operatorname{grad}_0 f$$

where $\mathbf{g} = \mathbf{h}_E$ is the physical space metric in the frame E, and where $\operatorname{grad}_\mathbf{g}$ (resp. $\operatorname{grad}_0$) is the gradient operator relative to metric $\mathbf{g}$ (resp. relative to the "natural" metric $\mathbf{g}^0$, with constant curvature, of which the "ether" network (3-D manifold) M=N$_E$ is assumed to be equipped with). And the line element of the space-time metric $\boldsymbol{\gamma}$, affected by gravitational contraction of the space standards (relative to metric **g**$^0$) and by gravitational dilation of the time standards (relative to the "absolute time" $T$), has the form

(2.20) $\qquad ds^2 = \beta^2 (dx^0)^2 - dl^2, \qquad x^0 = cT,$

where $dl^2$ is the line element of metric **g**. This has the following simple expression in "isopotential" coordinates ($y^\alpha$), i.e. coordinates such that, at a given time $T$, $y^1$ = Const (in space) is equivalent to $p_e$ = Const, and that the natural metric **g**$^0$ is diagonal, $(g^0_{ij})$ = diag $(a^0_i)$:

(2.21) $\qquad (g_{ij}) = \text{diag}(a_i) \quad \text{with} \quad a_1 = a^0_1/f, \quad a_2 = a^0_2, \quad a_3 = a^0_3.$

For a time-dependent field $p_e$, such coordinates are *not* bound to the ether frame [4]. From Eq. (2.20), it follows that, if one selects any coordinates ($x^\alpha$), with $x^0 = cT$, that *are* bound to the frame E, then the components $\gamma_{0i}$ are zero. Thus a simultaneity is defined for the frame E as a whole, in other words the absolute time $T$ is a universal time in the frame E. For the important case of an insular matter distribution, the absolute time $T$ is the local time measured at any point $\mathbf{x}_0$ which is far enough from matter so that no gravitation field is felt there. Moreover, the global synchronization condition ($\gamma_{0i} = 0$) does not hold true in a frame that rotates rigidly with respect to E, nor in general in a frame that moves uniformly with respect to E [3] (the condition $\gamma_{0i} = 0$ holds true for any frame in uniform translation, in the case that no gravitation field is present, thus for the flat metric $\boldsymbol{\gamma} = \boldsymbol{\gamma}^0$). These considerations justify the denomination "absolute time" for $T$. Hence, the ether frame E, which is already a global inertial frame in the sense that the mass force in E (2.18-19) is purely gravitational, is really a physically privileged reference frame (according to this theory).

## 3. Extended Newton law and geodesic motion
### *3.1 A possible form for the gravitation force in a globally synchronized reference frame*

We now investigate the possible form of the gravitation force. In order to make some meaningful induction from Newtonian theory, it is very useful to work in a reference frame F, in which the $\gamma_{0i}$ components of metric $\boldsymbol{\gamma}$ are zero (Subsect. 2.1). The concept of global simultaneity is indeed so deeply involved in any Newtonian analysis, that any induction from Newtonian theory to the general situation with curved space-time, where a simultaneity is defined only along a trajectory, would seem dangerous. Whereas, if one works in a frame such that $\gamma_{0i} = 0$, the only change in the time concept is that now the clocks go differently at different positions and times [Eq. (2.4)]. We note that the existence of a frame F, in which the $\gamma_{0i}$'s are zero, is not a physically restrictive assumption, since it breaks down only for rather pathological space-times: in "normal" space-times it is even possible to select a "synchroneous" frame which not only enjoy this global synchronization, but in which the $\gamma_{00}$ component is uniform, i.e. the local time flows uniformly (Landau & Lifchitz [11], Mavrides [16]). Thus it "normally" exists *many* different frames such that $\gamma_{0i} = 0$. Which form of the gravitation force could one consistently state in such a reference frame?

---

[3] Here, rigid rotation and uniform motion *can* be defined, at least if the metric manifold (M, **g**$^0$) has *zero* curvature, i.e. is Euclidean.

For the class of theories considered in Section 2, what is considered by any such theory as "the gravitation field" has been assumed to determine the space-time metric **γ** (for non-covariant theories, we should add that this has only to be true in some preferred reference frame which is like E, i.e. such that $\gamma_{0i} = 0$). Here, we will assume, in a more restrictive way, that *the metric field **γ** contains the gravitation field* (at least in the preferred frame). This is true in any reference frame for GR and for the "relativistic theory of gravitation" (RTG) proposed by Logunov *et al.* [13-14], and this is true in the ether frame E in the tentatively proposed theory. On the other hand, in order that SR holds true locally and that the inertial and (passive) gravitational mass coincide, the gravitation force must have the form

(3.1) $$\mathbf{F}_g = m(v)\,\mathbf{g},$$

with **g** a space vector in the considered frame. If we want that the metric field plays the rôle of a potential, we must ask that **g** depends linearly on the first derivatives of **γ**, and bearing in mind Newtonian theory we should add that only the *spatial* derivatives $\gamma_{\mu\nu,k}$ are allowed. But, in a frame where $\gamma_{0i} = 0$, we have $\gamma_{ij} = -h_{ij}$ with **h** the space metric in this frame, i.e. the metric **γ** reduces to the joint data **γ** = (*f*, **h**) with $f = \gamma_{00}$. Thus, we are looking for a space vector **g** depending linearly on the spatial derivatives of *f* and **h**. To be contravariant by general space transformation, **g** must depend linearly on the *covariant derivatives* of *f* and **h** (with respect to the space metric **h**!). But, as is known, the covariant derivatives of metric **h** with respect to **h** itself are all zero (in other words, one may cancel all spatial derivatives $h_{ij,k}$ at any given point by a purely spatial coordinate transformation). Hence, **g** should have the form

(3.2) $$\mathbf{g} = a(f, \mathbf{h})\, \mathrm{grad}_\mathbf{h}\, f,$$

where $a$ must be a given function of the *values* of the metric fields at the considered point $(x^\alpha)$ in space-time, $f = f(x^\alpha)$ and $\mathbf{h} = \mathbf{h}(x^\alpha)$ in Eq. (3.20), thus $a(f, \mathbf{h})$ is completely independent of the *variation* of $f$ and **h** with time and position.

Now we add the condition that *geodesic motion (Einstein's assumption) must apply to free particles ($\mathbf{F}_0 = 0$) for a static gravitation field*. This is exactly equivalent to assuming the following expression for the gravitation force *in the static case*:

(3.3) $$\mathbf{F}_g = -m(v)\, c^2\, \frac{\mathrm{grad}_\mathbf{h}\, \beta}{\beta} = m(v)\, \mathrm{grad}_\mathbf{h}(-c^2 \mathrm{Log}\, \beta), \quad \text{where } \beta \equiv \sqrt{\gamma_{00}},$$

Indeed, it was already proved (and it will be proved again below, in a different way) that Eq. (3.3), which occurs naturally in the ether theory, implies geodesic motion for mass particles in the static case [2]; this is also true for photons [3], substituting in that case the mass content of the energy $e = h\nu$ for the inertial mass $m(v)$. Conversely it is proved in Landau & Lifchitz [11] that geodesic motion implies the expression (3.3) for the force in the static case, defined as the derivative (2.7) of the momentum (2.6) [4]. Thus the reason for assuming geodesic motion in the static case is that it is indeed so for the tentative ether theory as well as, of course (and in any situation) for usual theories of gravitation with curved space-time, in particular GR and the RTG. So we must have, by Eqs. (3.1), (3.2) and (3.3):

(3.4) $$\mathbf{g} = -c^2\, \frac{\mathrm{grad}_\mathbf{h}\, \beta}{\beta} = -\frac{c^2}{2}\, \frac{\mathrm{grad}_\mathbf{h}\, f}{f}, \quad \text{i.e. } a(f, \mathbf{h}) = -\frac{c^2}{2f},$$

when $f_{,0} = 0$ and $\mathbf{h}_{,0} = 0$. But since $a(f, \mathbf{h})$ depends only on the local values of $f$ and **h**, not on their variation, Eq. (3.2) implies then that **g** keeps the form (3.4) and thus Eq. (3.3) holds true in the most general situation.

---

[4] Actually, Landau & Lifchitz [11, § 88] derived from geodesic assumption the expression of the force in the *stationary* case, using this same definition for the force (as is consistent with the present work, Subsect. 2.2). They found an expression involving an additional term which cancels if $\gamma_{0i} = 0$.

*3.2 Expression of the 4-acceleration for a "free" particle using the extended Newton law*

In theories with a (pseudo-) Riemannian space-time metric, two well-known space-time vectors may be defined for a time-like test particle (i.e. a mass point). These are the 4-velocity **U**, which is the velocity on the world line of the particle in space-time, when the world line is parametrized with the proper time $\tau$ of the particle,

(3.5) $$U^\alpha \equiv dx^\alpha/d\tau,$$

and the 4-acceleration **A**, which is the absolute derivative $\Delta U/\Delta\tau$ of the former relative to the *space-time* metric **γ**. Thus

(3.6) $$A^\alpha \equiv \left(\frac{\Delta U}{\Delta \tau}\right)^\alpha \equiv \frac{dU^\alpha}{d\tau} + \Gamma'^\alpha_{\mu\nu} U^\mu \frac{dx^\nu}{d\tau} \equiv \frac{dU^\alpha}{d\tau} + \Gamma'^\alpha_{\mu\nu} U^\mu U^\nu,$$

the $\Gamma'^\alpha_{\mu\nu}$ 's being the Christoffel symbols of metric **γ** in coordinates $(x^\alpha)$.

(**i**) Spatial components of the 4-acceleration in a globally synchronized reference frame. It is recalled that we use coordinates $(x^\alpha)$ that are bound to a "globally synchronized" frame **F**. Thus $\gamma_{0i} = 0$ ($i = 1, 2, 3$), from which follows immediately that:

(3.7) $$h_{ij} = -\gamma_{ij}, \quad \Gamma^i_{jk} = \Gamma'^i_{jk},$$

hence

(3.8) $$\left(\frac{\Delta U}{\Delta \tau}\right)^i = \frac{dU^i}{d\tau} + \Gamma^i_{jk} U^j U^k + \Gamma'^i_{00}(U^0)^2 + 2\Gamma'^i_{0k} U^0 U^k.$$

In this equation, we note that, in view of Eq. (3.7)$_1$ (and since $h^{ij} = -\gamma^{ij}$ is always true):

(3.9) $$\Gamma'^i_{0k} U^0 U^k = \gamma^{i\alpha} \frac{(\gamma_{\alpha 0,k} + \gamma_{\alpha k,0} - \gamma_{0k,\alpha})}{2} U^0 U^k = \frac{1}{2} h^{ij} h_{jk,0} \frac{dx^0}{d\tau} U^k.$$

By (2.4) and (3.5) we get:

$$U^0 = (dx^0/dt_x)(dt_x/d\tau) = c\,(dt_x/d\tau)/\sqrt{\gamma_{00}},$$

but, using Eqs. (2.1)-(2.3) and (2.5), it may be proved (cf. Landau & Lifchitz [11]) that, independently of the fact that $\gamma_{0i} = 0$, one has always:

$$\frac{dt_\mathbf{x}}{d\tau} = \gamma_v, \tag{3.10}$$

as was already noted [2] for the tentative theory. Hence we obtain

$$U^0 = \frac{c\gamma_v}{\sqrt{\gamma_{00}}} = \frac{c\gamma_v}{\beta}, \tag{3.11}$$

so we reexpress another term in Eq. (3.8), calculating $\Gamma'^i_{00}$ as for $\Gamma'^i_{0k}$ in Eq. (3.9) and using again Eq (2.4):

$$\Gamma'^i_{00}(U^0)^2 = h^{ij}\frac{\gamma_{00,j}}{2}\frac{c^2\gamma_v^2}{\beta^2} = h^{ij}\frac{2\beta\,\beta_{,j}}{2}\frac{c^2\gamma_v^2}{\beta^2} = \gamma_v^2 c^2 \frac{(\mathrm{grad}_\mathbf{h}\beta)^i}{\beta}.$$

We recognize here the component $g^i$ of the assumed gravity acceleration [Eq. (3.4)], thus

$$\Gamma'^i_{00}(U^0)^2 = -\gamma_v^2 g^i. \tag{3.12}$$

It is now possible to calculate $(\Delta U/\Delta\tau)^i$ with the Newton law, for a "free" particle [Eq. (2.8) with $\mathbf{F}_0 = 0$ and with $\mathbf{F}_g$ given by Eq. (3.1)]. In a first step, let us calculate with the incompletely defined Newton law, which is obtained if one uses the derivative $D_\lambda \mathbf{P}/Dt_\mathbf{x}$ with the unspecified parameter $\lambda$ [cf. Eq. (2.14)]. Using (3.10), we may write this in terms of $\tau$:

$$(D_\lambda \mathbf{P}/Dt_\mathbf{x})^i \equiv (D_\lambda \mathbf{P}/D\tau)^i/\gamma_v = m_0\,\gamma_v\,g^i,$$

and we have by Eqs. (2.5), (2.6) and (3.10):

$$P^i = m_0\,\gamma_v\,v^i = m_0\,\gamma_v\,dx^i/dt_\mathbf{x} = m_0\,dx^i/d\tau = m_0\,U^i, \tag{3.13}$$

so the "unspecified" Newton law writes

$$(D_\lambda \mathbf{u'}/D\tau)^i = \gamma_v^2\,g^i, \tag{3.14}$$

where $\mathbf{u}' \equiv (U^i)$ means the spatial part of the 4-velocity $\mathbf{U}$. Applying definition (2.14) which involves terms given by Eqs. (2.13) and (2.7), we get

$$(3.15) \qquad \left(\frac{D_\lambda \mathbf{u}'}{D\tau}\right)^i = \frac{dU^i}{d\tau} + \Gamma^i{}_{jk} U^j U^k + \lambda h^{ij} h_{jk,0} \frac{dx^0}{d\tau} U^k.$$

Hence, the unspecified Newton law imposes the following values to the spatial components (in coordinates bound to a globally synchronized frame F ) of the 4-acceleration of a free test particle [Eq. (3.8) with (3.9) and (3.12)], depending on the parameter $\lambda$:

$$(3.16) \qquad \left(\frac{\Delta \mathbf{U}}{\Delta\tau}\right)^i = 2(1-\lambda)\Gamma'^i{}_{0k} U^0 U^k = (1-\lambda)\ h^{ij} h_{jk,0} \frac{dx^0}{d\tau} U^k.$$

In particular, *the spatial part of the equation for space-time geodesics is satisfied for a variable gravitation field ($h_{jk,0} \neq 0$) if and only if the parameter $\lambda$ has the value $\lambda = 1$.*

(**ii**) Time component of the 4-acceleration in a globally synchronized frame

For the time component, we have simply

$$(3.17) \qquad A^0 = \left(\frac{\Delta \mathbf{U}}{\Delta\tau}\right)^0 \equiv \frac{dU^0}{d\tau} + \Gamma'^0{}_{00}(U^0)^2 + 2\Gamma'^0{}_{0k} U^0 U^k + \Gamma'^0{}_{ij} U^i U^j.$$

Using Eq. (3.7)$_1$ and the fact that $\gamma_{00} = \beta^2$ [Eq. (2.4)], the $\Gamma'^0{}_{\mu\nu}$'s are easily calculated:

$$\Gamma'^0{}_{00} = \frac{\beta_{,0}}{\beta}, \quad \Gamma'^0{}_{0k} = \frac{\beta_{,k}}{\beta}, \quad \Gamma'^0{}_{ij} = \frac{h_{ij,0}}{2\beta^2}.$$

By Eq. (3.11), which implies also that $U^k = (\gamma_v/\beta)(dx^k/dt)$, one then rewrites (3.17) as

$$(3.18) \qquad A^0 \frac{\beta}{c\gamma_v} = \frac{d}{dt}\left(\frac{\gamma_v}{\beta}\right) + \frac{\gamma_v}{\beta^2}\frac{\partial\beta}{\partial t} + 2\frac{\gamma_v}{\beta^2}\beta_{,k}\frac{dx^k}{dt} + \frac{1}{2c^2\beta^2}\frac{\gamma_v}{\beta}\frac{\partial h_{ij}}{\partial t}\frac{dx^i}{dt}\frac{dx^j}{dt}.$$

At this point, we may insert the energy balance deduced from the "unspecified" Newton law for the free test particle (Eq. (4.21) in ref. [4]):

$$\text{(3.19)} \qquad \frac{d}{dt}(\beta\gamma_v) = \gamma_v \frac{\partial \beta}{\partial t} + \beta\gamma_v \frac{1-2\lambda}{2c^2} \frac{\partial \mathbf{h}}{\partial t}(\mathbf{v},\mathbf{v})$$

with $v^i = (dx^i/dt)/\beta$ by Eqs. (2.4) and (2.5). [5] We have thus in Eq. (3.18):

$$\frac{d}{dt}\left(\frac{\gamma_v}{\beta}\right) = \frac{d}{dt}\left(\frac{1}{\beta^2}\beta\gamma_v\right) = \frac{1}{\beta^2}\frac{d}{dt}(\beta\gamma_v) + \beta\gamma_v\left[\frac{\partial}{\partial t}\left(\frac{1}{\beta^2}\right) + \left(\frac{1}{\beta^2}\right)_{,k}\frac{dx^k}{dt}\right]$$

$$= -\frac{\gamma_v}{\beta^2}\left(\frac{\partial \beta}{\partial t} + 2\beta_{,k}\frac{dx^k}{dt}\right) + \frac{\gamma_v}{\beta^3}\frac{1-2\lambda}{2c^2}\frac{\partial h_{ij}}{\partial t}\frac{dx^i}{dt}\frac{dx^j}{dt},$$

so that some cancellation occurs in (3.18). We obtain finally:

$$\text{(3.20)} \qquad A^0 = \frac{\gamma_v^2}{c\beta^4}(1-\lambda)\frac{\partial h_{ij}}{\partial t}\frac{dx^i}{dt}\frac{dx^j}{dt} = \frac{(1-\lambda)}{\beta^2}h_{ij,0}U^iU^j = 2(1-\lambda)\Gamma'^0_{ij}U^iU^j.$$

In particular, the time part of the equation for space-time geodesics, as well as the spatial part, is satisfied for a variable gravitation field ($h_{ij,0} \neq 0$) if and only if the parameter $\lambda$ has the value $\lambda = 1$. However, it is recalled that the value $\lambda = 1$ specifies the Newton law in an incorrect manner, since it means that Newton's second law is based on a vector time derivative which does not obey Leibniz' derivation rule.

Let us summarize the results of Subsects. 3.1 and 3.2, which concern Newton's second law and geodesic motion:

(NGM1) *Consider a theory with curved space-time metric* **γ** *and locally valid SR, and assume that in some "globally synchronized" reference frame* F *($\gamma_{0i} = 0$), the gravitation force (3.1) involves a space vector* **g** *depending only on the metric field* **γ**. *More precisely, assume that* **g** *does not depend on the time variation of* **γ** *and is linear with respect to the space variation of* **γ**. *In order that free particles follow space-time geodesics in the static case ($\gamma_{\mu\nu,0} = 0$), it is necessary and sufficient that the general expression of vector* **g** *in the frame* F *be*

---

[5] Eq. (3.19) is derived using the fact that *some* derivation rule of a scalar product can be obtained even with the "unspecified" Newton law, although it does not obey the true Leibniz rule [Eq. (2.9)] unless $\lambda = 1/2$. However, if $\lambda \neq 1/2$, this balance equation cannot be rewritten as a true conservation equation, at least in the scalar theory [1-4].

(3.21) $$\mathbf{g} = -c^2 \frac{\mathrm{grad}_\mathbf{h}\, \beta}{\beta} = -\frac{c^2}{2}\frac{\mathrm{grad}_\mathbf{h}\, f}{f}, \quad f \equiv \gamma_{00} \equiv \beta^2,$$

with **h** *the space metric in* **F**. *This expression implies Eqs. (3.16) and (3.20) for the 4-acceleration, thus it implies that, for a time-dependent field, geodesic motion corresponds exactly to the incorrect Newton law ($\lambda = 1$).*

### 3.3 *Characteristic form of the gravitation force associated with geodesic motion*

The assumption that the metric field $\gamma$ plays the rôle of a potential for the gravity acceleration **g** seems quite natural, if one thinks of a "soft" generalization of Newtonian gravity. The foregoing result implies, among other things, that Einstein's assumption of a motion following space-time geodesics is not such soft extension. But, after all, in Maxwell's theory the electric field involves also time derivatives of the electromagnetic potential, besides the usual space derivatives. Moreover, the Lorentz force depends on the velocity of the charged particle. A more general expression than we assumed for the gravity acceleration might hence be correct also, the more so as we now have empirical reasons to think that the gravity interaction indeed propagates, as does the electromagnetic field, and with the same velocity (Taylor & Weisberg [25]). That gravitation propagates with the velocity of light was first envisaged by Poincaré in his "electromagnetic", Lorentz-invariant theory of gravitation [20-21] and, as is well known, it is predicted by Einstein's theory.

Thus we now investigate the possible form of the vector **g**, subjected to the unique constraint that *geodesic motion should occur with the correct form of Newton's second law, i.e. $\lambda = 1/2$*. We continue to work in a globally synchronized reference frame and, in order to simplify the expressions, we take **g** in the form

(3.22) $$\mathbf{g} = -c^2 \frac{\mathrm{grad}_\mathbf{h}\, \beta}{\beta} + \mathbf{g'} = -\frac{c^2}{2}\frac{\mathrm{grad}_\mathbf{h}\, f}{f} + \mathbf{g'}, \quad f \equiv \gamma_{00} \equiv \beta^2.$$

Starting from Eq. (3.6) as before, nothing changes until Eq. (3.12), which now becomes

(3.23) $$\Gamma'^i_{00}(U^0)^2 = -\gamma_v^2\left(g^i - g'^i\right).$$

And again nothing changes until Eq. (3.16), which is modified into

$$(3.24) \quad A^i = \left(\frac{\Delta \mathbf{U}}{\Delta \tau}\right)^i = (1-\lambda)\, h^{ij} h_{jk,0} \frac{dx^0}{d\tau} U^k + \gamma_v^2 g'^i,$$

Hence, the spatial components of the 4-acceleration cancel with $\lambda = 1/2$, if and only if

$$(3.25) \quad g'^i = \frac{-1}{2\beta} h^{ij} \frac{\partial h_{jk}}{\partial t} v^k, \quad \text{i.e.} \quad \mathbf{g'} = \frac{-1}{2\beta} \mathbf{h}^{-1} \cdot \frac{\partial \mathbf{h}}{\partial t} \cdot \mathbf{v} = \frac{-1}{2} \mathbf{h}^{-1} \cdot \frac{\partial \mathbf{h}}{\partial t_{\mathbf{x}}} \cdot \mathbf{v}.$$

But does this expression also cancel the time part of the 4-acceleration? To check this, one must reexamine the energy balance derived in ref. [4]. Proceeding in the same way, we find easily that the energy balance resulting from the expression (3.22, 3.25) of $\mathbf{g}$ is (with $\lambda = 1/2$)

$$(3.26) \quad \frac{d}{dt}(\beta \gamma_v) = \gamma_v \frac{\partial \beta}{\partial t} - \frac{\beta \gamma_v}{2c^2} \frac{\partial \mathbf{h}}{\partial t}(\mathbf{v}, \mathbf{v}),$$

instead of Eq. (3.19). Thus, with the correct Newton law ($\lambda = 1/2$), the same expression is now obtained as it was obtained before with the incorrect Newton law ($\lambda = 1$). Therefore, the time part of the geodesic equation, $A^0 = 0$, is satisfied for $\lambda = 1/2$, as it was previously for $\lambda = 1$. We have proved the following:

(NGM2) *Consider a theory with curved space-time metric* $\gamma$ *and locally valid SR, and assume the correct time derivative (2.15) in the extension (2.8) of Newton's second law. In order that free particles [$\mathbf{F}_0 = 0$ in Eq. (2.8)] follow space-time geodesics, it is necessary and sufficient that, in any globally synchronized reference frame* F *($\gamma_{0i} = 0$), the gravitation force (3.1) involve the following expression for the gravity acceleration (space vector* $\mathbf{g}$ *) :*

$$(3.27) \quad \mathbf{g}_{\text{geod}} = -c^2 \frac{\text{grad}_{\mathbf{h}} \beta}{\beta} - \frac{1}{2\beta} \mathbf{h}^{-1} \cdot \frac{\partial \mathbf{h}}{\partial t} \cdot \mathbf{v}, \quad \beta \equiv \sqrt{\gamma_{00}},$$

with $\mathbf{h}$ *the space metric in* F *and* $\mathbf{v}$ *the velocity vector [Eq. (2.5)].*

This result provides the general link between Newton's second law and Einstein's geodesic assumption.

### 4. Comparison with the literature

4.1 *Møller's work and the relation between covariant and contravariant form of Newton's law*

Among attempts to define Newton's second law in the case of a variable gravitation field, a well-known one is that of Møller [18]. However, Møller uses the absolute derivative with respect to the "frost" space metric, thus $\lambda = 0$ in Eq. (2.14), so that Leibniz' rule is not satisfied with the actual, time-dependent metric. In connection with this, he notes that this derivative does not commute with raising or lowering the indices with respect to the space metric **h**. As a consequence, when he rewrites the equations for space-time geodesics in the form of Newton's second law with gravitational forces, the latter look very different in covariant and in contravariant form. We show that this difficulty is absent with our definition.

Indeed, it is easy to adapt our line of reasoning so as to define the time-derivative of a spatial *covector* **w**\*. One finds in exactly the same way that, apart from Leibniz' rule, a one-parameter family of time-derivatives may defined as:

(4.1) $\qquad D_\lambda \mathbf{w}^*/D\chi \equiv D_0 \mathbf{w}^*/D\chi - \lambda\, \mathbf{t}.\mathbf{w}^*$,

with

(4.2) $(\mathbf{t}.\mathbf{w}^*)_i \equiv h_{ij,0}(dx^0/d\chi)h^{jk}w^*_k \equiv (dx^0/d\chi)(\mathbf{h}_{,0}.\mathbf{h}^{-1})_i{}^k w^*_k$

$\qquad\qquad\qquad = (dx^0/d\chi)(\mathbf{h}^{-1}.\mathbf{h}_{,0})^k{}_i w^*_k = t^k{}_i w^*_k$,

and where $D_0 \mathbf{w}^*/D\chi$ is the absolute derivative using the "frost" metric. And one finds that Leibniz' rule imposes $\lambda = 1/2$. It is also easy to verify that, for this correct value $\lambda = 1/2$ and, for a time-dependent metric **h**, *only* for this value, the time-derivative $D_\lambda /D\chi$ *does commute* with raising or lowering the indices with respect to the space metric **h**, that is

(4.3) $\qquad D_{1/2}(\mathbf{h}.\mathbf{w})/D\chi = \mathbf{h}.(D_{1/2}\mathbf{w}/D\chi)$.

Therefore, if one takes the covariant components of the momentum instead of the contravariant ones, thus substituting $\mathbf{P}^* = \mathbf{h}.\mathbf{P}$ for **P**, then the corresponding "covariant Newton law" will involve just the covariant components of the force, $\mathbf{F}^* = \mathbf{h}.\mathbf{F} = \mathbf{h}.(\mathbf{F}_0 + \mathbf{F}_g)$ in Eq. (2.8).

4.2 *Newton's second law with the "Fermi-Walker" time-derivative*

From now on, we will discuss the work on "Newton's second law in relativistic gravity" as reviewed and unified by Jantzen *et al*. [10]. They define the equivalent of what we call a frame (spatial network) by a 4-velocity vector field **u**, and they name it "observer congruence". What they call "observer-adapted frames" is a very different notion from that of adapted coordinates as defined by Møller [18] and Cattaneo [6-7]. Here we continue to work in adapted coordinates, i.e. such that the observers of the network (or congruence) have constant space coordinates. In such coordinates, the contravariant and covariant components of **u** are given by

$$(4.4) \qquad (u^\alpha) = \left(\frac{1}{\sqrt{-\gamma_{00}}}, 0, 0, 0\right), \quad (u_\alpha) = \left(-\sqrt{-\gamma_{00}}, \left(\frac{\gamma_{0i}}{\sqrt{-\gamma_{00}}}\right)_{i=1,2,3}\right)$$

(we keep our notations, except for the fact that we set $u^\alpha \equiv dx^\alpha/ds$ and adopt the (3,1) signature as in refs. [6-7] and [10], until the end of this Section). It follows that the spatial projection tensor $\Pi = \Pi(\mathbf{u})$ [7,10], which is a space-time tensor defined in general by

$$\Pi^\mu{}_\nu \equiv \delta^\mu{}_\nu + u^\mu u_\nu,$$

has a simple expression:

$$(4.5) \qquad \Pi^i{}_j = \delta^i{}_j, \quad \Pi^i{}_0 = 0, \quad \Pi^0{}_j = -\gamma_{0j}/\gamma_{00}, \quad \Pi^0{}_0 = 0.$$

It corresponds to the projection of the local tangent space to space-time onto the hyperplane which is **γ**-perpendicular to the local 4-velocity **u** of the observer congruence. In connection with this, what is called a "spatial tensor" by Cattaneo [7] and by Jantzen et al. [10] is also a very different notion from that used by Møller [18] and in the rest of this paper. For us (and for Møller), a spatial tensor is just an element of a tensor space at the relevant point of the spatial network (3-D Riemannian manifold) N, thus its components depend on the three spatial (Latine) indices only, $i = 1, 2, 3$, in adapted coordinates. In refs. [7], [10] and in the remainder of this Section, a spatial tensor is a *space-time tensor* which is *equal to its projection*, the latter being generally defined by Eq. (2.2) of ref. [10]. E.g. for a 4-vector (space-time vector) **X**, the projection writes:

$$(4.6) \qquad (\mathbf{\Pi}.\mathbf{X})^\alpha = \Pi^\alpha{}_\mu X^\mu.$$

Hence in adapted coordinates, by (4.5):

(4.7) $$(\Pi.\mathbf{X})^i = X^i, \quad (\Pi.\mathbf{X})^0 = -\gamma_{0j}X^j/\gamma_{00},$$

so that the "time" component $X^0$ is *not* equal to zero for a "spatial vector" (except for a "normal congruence", i.e. the case where $\gamma_{0j} = 0$ in adapted coordinates). We note also that the "rescaled time" $\tau_{(\mathbf{U},\mathbf{u})}$ considered in ref. [10] (for a time-like test particle with 4-velocity **U**), as well as the "standard time" $T$ considered in refs. [6-7], is the same variable as our "local time" $t_\mathbf{x}$, synchronized along the trajectory of the test particle, with their $\gamma = \gamma_{(\mathbf{U},\mathbf{u})}$ being our $\gamma_v$ [Eqs. (2.3) and (3.10) here]. On the other hand, what is called the "Fermi-Walker total spatial covariant derivative" (fw TSCD) in ref. [10], has the following expression for an arbitrary parameter $\chi$ (although it is defined only for $\chi = \tau_{(\mathbf{U},\mathbf{u})} \equiv t_\mathbf{x}$ in ref. [10]):

(4.8) $$\frac{D_{(\text{fw})}\mathbf{X}}{D\chi} \equiv \Pi.\frac{\Delta \mathbf{X}}{\Delta \chi}.$$

We have thus in adapted coordinates, by Eq. (4.7):

(4.9) $$\left(\frac{D_{(\text{fw})}\mathbf{X}}{D\chi}\right)^i \equiv \left(\frac{\Delta \mathbf{X}}{\Delta \chi}\right)^i \equiv \left(\frac{dX^i}{d\chi} + \Gamma'^i_{\mu\nu}X^\mu \frac{dx^\nu}{d\chi}\right), \quad i = 1,2,3,$$

and the "time" part of the derivative is not independent from the "space" part:

(4.10) $$\left(\frac{D_{(\text{fw})}\mathbf{X}}{D\chi}\right)^0 \equiv -\frac{\gamma_{0j}}{\gamma_{00}}\left(\frac{D_{(\text{fw})}\mathbf{X}}{D\chi}\right)^j.$$

What corresponds to Newton's second law in [10] is the evaluation of the spatial projection of the 4-acceleration **A** of the test particle. Apart from the different notation, it amounts almost exactly to Eq. (2.8) here, with the same definition (2.6) for the momentum, involving the same relative velocity (2.5), though with the derivative defined by Eq. (4.8) instead of Eq. (2.15). One difference is that the velocity **v** and momentum **P** are now spatial 4-vectors which turn out to be the respective projections of the 4-vectors **U'** and **P'**, with **U'** the 4-velocity **U**, rescaled to the local time, and **P'** the usual 4-momentum. Thus the spatial components of **v** and **P** are the same as in this work, and the "time" components obey the general rule for a spatial vector **X**, i.e. such that $\Pi.\mathbf{X} = \mathbf{X}$:

(4.11) $$X^0 = -\gamma_{0j}X^j/\gamma_{00}.$$

Another difference is that the gravitational force, which is the total force for a free particle, is hence deduced, in the frame of GR and other "metric theories", from the geodesic equation, i.e. **A** = 0.

Having thus recognized that the spatial part (4.9) of the derivative (4.8) plays exactly the same rôle in ref. [10] as the derivative (2.15) plays here, we may comment on the difference between the two derivatives. Since the spatial components (4.9) are just those of the space-time absolute derivative $\Delta \mathbf{X}/\Delta \chi$, the Fermi-Walker TSCD involves space-time coupling in a generally inextricable way, in that it cannot in general be defined in terms of only the spatial metric **h** and the local time $t_x$. Hence, this derivative cannot be used in an arbitrary reference frame to define a "true" Newton law as it has been defined here, i.e. precisely a law involving only the separate space and time metrics in the given reference frame, thus allowing to "forget" the concept of space-time as long as one does not change the reference frame.

4.3 *The "normal" and "corotational" Fermi-Walker derivatives obey Leibniz' rule*

Surprisingly, the question whether the introduced time-derivatives satisfy the Leibniz rule is not investigated in refs. [6, 7, 10]. However, it is not difficult to show that *the two Fermi-Walker derivatives do verify Eq. (2.9)*. The spatial metric in those works is of course the same thing as here, except for the indices and the signature:

(4.12) $$h_{\alpha\beta} \equiv \gamma_{\alpha\beta} + u_\alpha u_\beta \Rightarrow h_{ij} \equiv \gamma_{ij} - \frac{\gamma_{0i}\gamma_{0j}}{\gamma_{00}}, \ h_{0i} = h_{i0} = h_{00} = 0.$$

Using Eqs. (4.12) and (4.11), one verifies that, for two *spatial* space-time vectors **X** and **Y**:

(4.13) $$\mathbf{h}(\mathbf{X},\mathbf{Y}) = \boldsymbol{\gamma}(\mathbf{X},\mathbf{Y}).$$

On the other hand, the absolute space-time derivative obeys the Leibniz rule:

(4.14) $$\frac{d}{d\chi}[\boldsymbol{\gamma}(\mathbf{X},\mathbf{Y})] = \boldsymbol{\gamma}\left(\mathbf{X},\frac{\Delta \mathbf{Y}}{\Delta \chi}\right) + \boldsymbol{\gamma}\left(\frac{\Delta \mathbf{X}}{\Delta \chi},\mathbf{Y}\right).$$

Since the "normal" Fermi-Walker derivative is a spatial vector whose spatial components are just those of the absolute space-time derivative [Eqs. (4.8) and (4.9)], we may use Eq. (4.13) to rewrite Eq. (4.14) as:

$$(4.15) \qquad \frac{d}{d\chi}[\mathbf{h}(\mathbf{X},\mathbf{Y})] = \mathbf{h}\left(\mathbf{X}, \frac{D_{(\text{fw})}\mathbf{Y}}{D\chi}\right) + \mathbf{h}\left(\frac{D_{(\text{fw})}\mathbf{X}}{D\chi}, \mathbf{Y}\right).$$

This is the Leibniz rule for the two spatial vectors $\mathbf{X}$ and $\mathbf{Y}$.

The "corotational" Fermi-Walker (cfw) derivative is related to the "normal" Fermi-Walker derivative by [10]:

$$(4.16) \qquad \left(\frac{D_{(\text{cfw})}\mathbf{X}}{D\chi}\right)^\alpha = \left(\frac{D_{(\text{fw})}\mathbf{X}}{D\chi}\right)^\alpha + \omega^\alpha{}_\mu \frac{c\, dt_\mathbf{x}}{d\chi} X^\mu.$$

Here $\omega^\alpha{}_\mu$ are the mixed components of the "spin-rate" space-time tensor. This comes from the decomposition of the covariant "spatial 4-velocity gradient",

$$(4.17) \quad \mathbf{k} = \mathbf{k}(\mathbf{u}) = -\Pi . \nabla^{(\gamma)} u^b, \qquad u^b \equiv (u_\alpha), \qquad k_{\alpha\beta} = -\Pi^\lambda{}_\alpha \Pi^\mu{}_\beta\, u_{\lambda;\mu},$$

into symmetric and antisymmetric part:

$$(4.18) \qquad k_{\alpha\beta} = -\theta_{\alpha\beta} + \omega_{\alpha\beta}, \quad -\theta_{\alpha\beta} = (k_{\alpha\beta} + k_{\beta\alpha})/2, \quad \omega_{\alpha\beta} = (k_{\alpha\beta} - k_{\beta\alpha})/2,$$

and the mixed components $\omega^\alpha{}_\mu$ are obtained by raising the index $\alpha$ with metric $\gamma$. It appears that, just like the ordinary one, the corotational Fermi-Walker derivative cannot in general be expressed in terms of the spatial metric $\mathbf{h}$ and the local time $t_\mathbf{x}$ only. Moreover, it is difficult here to refrain from asking the question: with respect to *what* does the "spin rate" $\omega$ measure the rate of relative spin of the considered reference fluid (network)? Already the understanding of the *strain* rate $\theta$ is difficult: without any preferred reference fluid, we may only define, so to speak, the "strain rate of the fluid with respect to itself" due to the evolution of the spatial metric $\mathbf{h}$, and this is precisely what measures the $\mathbf{t} = \mathbf{h}^{-1}.\mathbf{h}_{,0}\,(dx^0/dt_\mathbf{x})$ tensor in our derivative (2.15) (with $\chi = t_\mathbf{x}$) - but the tensors $\mathbf{t}$ and $\theta$ are two different objects.

As to Leibniz' rule, it applies as for the ordinary fw derivative. Indeed, due to the antisymmetry of the covariant tensor $\omega$ [Eq. (4.18)$_3$], the definition (4.16) gives

$$\gamma\left(\mathbf{X}, \frac{D_{(\mathrm{cfw})}\mathbf{Y}}{D\chi}\right) + \gamma\left(\frac{D_{(\mathrm{cfw})}\mathbf{X}}{D\chi}, \mathbf{Y}\right) - \gamma\left(\mathbf{X}, \frac{D_{(\mathrm{fw})}\mathbf{Y}}{D\chi}\right) - \gamma\left(\frac{D_{(\mathrm{fw})}\mathbf{X}}{D\chi}, \mathbf{Y}\right) =$$

$$= \frac{c\,dt_{\mathbf{x}}}{d\chi} \gamma_{\mu\nu}\left(\omega^{\mu}{}_{\rho} X^{\rho} Y^{\nu} + \omega^{\nu}{}_{\rho} Y^{\rho} X^{\mu}\right) = \frac{c\,dt_{\mathbf{x}}}{d\chi}\left(\omega_{\nu\rho} X^{\rho} Y^{\nu} + \omega_{\mu\rho} Y^{\rho} X^{\mu}\right) = 0.$$

The Leibniz rule follows from this by (4.13) and (4.15), the cfw derivative being also a spatial vector [10]:

(4.19) $\mathbf{h}\left(\mathbf{X}, \frac{D_{(\mathrm{cfw})}\mathbf{Y}}{D\chi}\right) + \mathbf{h}\left(\frac{D_{(\mathrm{cfw})}\mathbf{X}}{D\chi}, \mathbf{Y}\right) = \mathbf{h}\left(\mathbf{X}, \frac{D_{(\mathrm{fw})}\mathbf{Y}}{D\chi}\right) + \mathbf{h}\left(\frac{D_{(\mathrm{fw})}\mathbf{X}}{D\chi}, \mathbf{Y}\right) = \frac{d}{d\chi}[\mathbf{h}(\mathbf{X}, \mathbf{Y})].$

4.4 *The case of a globally synchronized frame and the "Lie" time-derivative*

We consider the particular case of a *globally synchronized frame* (or "normal congruence"), in which the $\gamma_{0i}$ components of the space-time metric are zero in some adapted coordinates. Then the spatial projection tensor $\Pi$ [Eq. (4.5)] writes simply

(4.20) $$(\Pi_{\nu}{}^{\mu}) = \mathrm{diag}(0,1,1,1)$$

in such coordinates. Hence, in such coordinates, substituting its spatial projection $\Pi(\mathbf{u}).\mathbf{T}$ for a space-time tensor $\mathbf{T}$ amounts exactly to taking its space components only. In particular, the "time" component of a spatial vector $\mathbf{X}$ is now equal to zero. Moreover, the spatial Christoffel symbols of the space-time metric are equal to the Christoffel symbols of the spatial metric [Eq. (3.7)]. This implies that the Fermi-Walker derivative coincides, for the case considered and for a *spatial* vector $\mathbf{X}$ (thus $X^0 = 0$), with the $D_{1/2}$ derivative. Indeed, using Eq. (3.9), we find:

(4.21) $\left(\frac{D_{(\mathrm{fw})}\mathbf{X}}{D\chi}\right)^i \equiv \left(\frac{\Delta \mathbf{X}}{\Delta \chi}\right)^i \equiv \frac{dX^i}{d\chi} + \Gamma^i_{jk} X^j \frac{dx^k}{d\chi} + \Gamma'^i_{j0} X^j \frac{dx^0}{d\chi} =$

$$= \frac{dX^i}{d\chi} + \Gamma^i_{jk} X^j \frac{dx^k}{d\chi} + \frac{1}{2} h^{ik} h_{kj,0} X^j \frac{dx^0}{d\chi} = \left(\frac{D_{1/2}\mathbf{X'}}{D\chi}\right)^i,$$

with $\mathbf{X'} \equiv (X^i)$.

For the non-zero components of the **k** tensor [Eq. (4.17)], we obtain using Eqs. (4.20), (3.9) and (4.4) [and since $h_{jk} = \gamma_{jk}$ with the (3,1) signature]:

$$(4.22) \quad -k_{ij} = u_{i;j} = h_{ik} u^k{}_{;j} = h_{ik} \Gamma'^k{}_{0j} u^0 = \frac{1}{2} h_{ij,0} u^0 = \frac{1}{2} h_{ij,0} \frac{dx^0}{c\,dt_\mathbf{x}}$$

Therefore, the "spin-rate" tensor **ω** is nil for a normal congruence [6], so that the corotational Fermi-Walker derivative coincides, for spatial vectors, with the "normal" one, and thus with the proposed derivative, $D = D_{1/2}$. On the other hand, we have from (4.18) and (4.22):

$$\theta_{ij} = -k_{ij} = \frac{1}{2} h_{ij,0} \frac{dx^0}{c\,dt_\mathbf{x}} \ .$$

What is called "Lie" TSCD derivative in ref. [10], is not a Lie derivative in the usual sense but the projection of a Lie derivative [10], and is defined in general by [10]:

$$(4.23) \quad \left(\frac{D_{\text{(lie)}}\,\mathbf{X}}{D\chi}\right)^\alpha = \left(\frac{D_{\text{(fw)}}\mathbf{X}}{D\chi}\right)^\alpha + \frac{c\,dt_\mathbf{x}}{d\chi}\left(\omega^\alpha{}_\mu X^\mu - \theta^\alpha{}_\mu X^\mu\right)$$

(extending again the definition [10] to an arbitrary parameter $\chi$). Hence, we have here:

$$(4.24) \quad \left(\frac{D_{\text{(lie)}}\,\mathbf{X}}{D\chi}\right)^i = \left(\frac{D_{\text{(fw)}}\mathbf{X}}{D\chi}\right)^i - \frac{1}{2}\frac{c\,dt_\mathbf{x}}{d\chi} h^{ik} h_{kj,0} \frac{dx^0}{c\,dt_\mathbf{x}} X^j = \left(\frac{D_0 \mathbf{X}}{D\chi}\right)^i .$$

In other words, the so-called "Lie" derivative coincides in that case with the absolute derivative with respect to the "frost" spatial metric, and so does not obey Leibniz' rule.

## 5. Concluding remarks

1. From our bibliographical research, it would appear that it had not yet been proposed in the literature the consistent definition of the time-derivative of a vector moving along a trajectory in a manifold equipped with a metric field $\mathbf{h}_\chi$ (the *spatial* metric in a given reference frame) that changes with the parameter $\chi$ on the trajectory, as it is proposed here. Indeed, of the three different notions of frame-dependent time-derivatives that have been reviewed and unified by Jantzen *et al*. [10], the two first ones (the Fermi-Walker derivatives) involve the whole *space-* *time* metric in an unseparable way, while the so-called "Lie" derivative does not obey Leibniz' rule. In our opinion, this would mean that no consistent *and* natural extension of Newton's second law to the case of a variable gravitation field in a general reference frame (in a theory with curved space-time as envisaged here) had yet been proposed either. It seems as if, from the orthodox relativistic point of view, it would be considered *a priori* impossible to define Newton's second law "really as before" - because the absolute priority is to maintain consistency with the notion that the 4-dimensional space-time is the essential physical reality. However, it turns out that the two Fermi-Walker derivatives coincide with the proposed derivative in the important case of a globally synchronized frame (or normal congruence).

2. We find that there is one and *only one natural* extension of Newton's second law to any theory with curved space-time metric, in the most general situation. In particular, one may *uniquely* identify that gravity acceleration $\mathbf{g}_{geod}$ which is necessary to obey Einstein's assumption, i.e. to obtain geodesic motion for free test particles. In doing so, we did not merely rewrite the three "spatial" equations for space-time geodesics as the space-vector relation "force = time-derivative of momentum": we also proved that the latter relation implies the "time" equation of geodesics, and this does not seem to have been done in earlier attempts. This "geodesic" gravity acceleration $\mathbf{g}_{geod}$ depends on the reference frame, as is natural in a "relativistic" theory (since the acceleration is not Lorentz-invariant). It may seem more surprising that $\mathbf{g}_{geod}$ depends on the velocity of the particle [Eq. (3.27)]. However, this is also the case for the Lorentz force which a charged particle undergoes in an electromagnetic field. The striking difference is that the magnetic force does not work, whereas the velocity-dependent part of $\mathbf{g}_{geod}$ does work. In the investigated case of a normal congruence, it has the same form as the Newtonian inertial force that appears in a reference frame undergoing pure strain with respect to an inertial frame [1]. But here this "inertial" force comes from the straining of the reference frame "with respect to itself" (i.e. due to the fact that the spatial metric evolves with time) and it cannot in general be cancelled by changing the reference frame. Thus, theories with geodesic motion inherently do not allow global inertial frames, although such global inertial frames do appear in their Newtonian limit. We also note that any velocity dependence of the gravity acceleration, $\mathbf{g} = \mathbf{g}(\mathbf{x}, \mathbf{v})$, implies that the definition of the passive gravitational mass, i.e. $m_g \equiv \mathbf{F}_g/\mathbf{g}$ with $\mathbf{F}_g$ the gravitation force, becomes indissolubly mixed with that of the gravity acceleration itself: one may change $\mathbf{g}$ and $m_g$ to $\alpha \mathbf{g}$ and $m_g/\alpha$ respectively, with $\alpha$ any scalar function of the velocity (e.g. $\alpha = \gamma_v^{\ n}$ where *n* is any real number), so that $m_g$ is operationally defined up to the arbitrary function $\alpha$ only. Hence, although Newton's second law *can* be defined in a "curved space-time" after all, the statement "$m_g$ = inertial mass $m(v)$" still remains partly conventional. Indeed, the only testable statement is then the universality of the gravitation force (which is really a crucial point, of course).

3. The identity between inertial and gravitational mass would have a stronger meaning if $\mathbf{F}_g$ would depend only on the position of a given test particle. However, for the kind of theories considered here, this could be true only in some preferred reference frame (this is, of course, in contrast with the Galilean situation). To check this identity, one might then define $\mathbf{g}$ for particles at rest in the preferred reference frame, thus $\mathbf{g}(\mathbf{x}) \equiv \mathbf{F}_g(\mathbf{v} = 0)/m_0$, and check experimentally whether or not $m_g \equiv \mathbf{F}_g/\mathbf{g}$ has the same velocity dependence as the inertial mass $m(v)$. In the scalar ether theory which has been tentatively proposed [1-4], a vector $\mathbf{g}$ depending only on the position, Eq. (3.21), has been found to occur naturally- consistently with the notion that $\mathbf{g}$ should be determined by the local state of some substratum. Thus this theory predicts "strong identity" between inertial and gravitational mass and, in connection with this, geodesic motion does not hold true in the general case in this theory. If one were to modify this theory so as to obtain geodesic motion, one would have to postulate Eq. (3.27) instead of Eq. (3.21). Then, the modified $\mathbf{g}$-field would still be determined (in the preferred frame E) by the scalar field $p_e$ or $\beta$ (together with the particle velocity!) However, this would lead to the energy balance (3.26), which has been seen to be incompatible with the obtainment of a true conservation equation for the energy in this scalar theory [4]. On the other hand, it might happen that this theory predict unobserved post-Newtonian effects of absolute motion.

## ACKNOWLEDGEMENT

I am grateful to Prof. P. Guélin for his useful comments on the manuscript.

Laboratoire 'Sols, Solides, Structures' [associated with the Centre National de la Recherche Scientifique], Institut de Mécanique de Grenoble, B.P. 53 X, F-38041 Grenoble cedex, France.